\documentclass[useAMS,usenatbib]{mn2e}

\usepackage[psamsfonts]{amssymb}
\usepackage[utf8]{inputenc}
\usepackage[T1]{fontenc}
\usepackage[dvips]{graphicx}
\usepackage{amsmath,alltt}
\usepackage{multirow}
\usepackage{rotating}
\usepackage{lscape}

\newcommand{\msun}{\>{\rm M_{\odot}}}

\newcommand{\mcmo}{M_{\rm{CMO}}}
\newcommand{\mnsc}{M_{\rm{NSC}}}
\newcommand{\mbh}{M_{\rm{SMBH}}}
\newcommand{\msph}{M_{\rm{Sph}}}
\newcommand{\rsph}{R_{\rm{Sph}}}
\newcommand{\tdyn}{\tau_{\rm{dyn}}}
\newcommand{\tgal}{\tau_{\rm{gal}}}
\newcommand{\ea}{et al.}

\newcommand{\kms}{\>{\rm km}\,{\rm s}^{-1}}

\newcommand{\kpc}{\>{\rm kpc}}

\newcommand{\gyr}{\>{\rm Gyr}}

\newcommand{\bdm}{\begin{displaymath}}
\newcommand{\edm}{\end{displaymath}}
\newcommand{\beq}{\begin{equation}}
\newcommand{\eeq}{\end{equation}}
\newcommand{\bit}{\begin{itemize}}
\newcommand{\eit}{\end{itemize}}
\newcommand{\ben}{\begin{enumerate}}
\newcommand{\een}{\end{enumerate}}
\newcommand{\bfi}{\begin{figure}[htb]}
\newcommand{\bpfi}{\begin{figure}[p]}

\title[Nuclear Clusters and their Host Spheroids]{Nuclear Star Clusters and the Stellar Spheroids of their Host Galaxies}
\author[Leigh, B{\"o}ker \& Knigge]{Nathan Leigh$^{1}$\thanks{E-mail: nleigh@rssd.esa.int}, 
Torsten B\"oker$^{1}$, Christian Knigge$^{2}$\\
$^{1}$European Space Agency, Space Science Department, Keplerlaan 1,
2200 AG Noordwijk, The Netherlands \\
$^{2}$School of Physics and Astronomy, University of Southampton,
Highfield, Southampton, SO17 1BJ, United Kingdom}
\begin{document}

\pagerange{\pageref{firstpage}--\pageref{lastpage}} \pubyear{2012}

\maketitle

\label{firstpage}

\begin{abstract}
We combine published photometry for the nuclear star clusters and stellar spheroids of 51 
low-mass, early-type galaxies in the Virgo cluster with empirical mass-to-light ratios, in order 
to complement previous studies that explore the dependence of nuclear star cluster (NSC) masses 
on various properties of 
their host galaxies.  We confirm a roughly linear relationship between NSC mass and luminous host 
spheroid mass, albeit with considerable scatter (0.57 dex). In order to translate this to an 
$\mnsc - \sigma$ relation, we estimate velocity dispersions from the virial theorem, assuming all 
galaxies in our sample share a common dark matter fraction and are dynamically relaxed. We then 
find that $\mnsc$ $\sim$ $\sigma^{2.73\pm 0.29}$, with a slightly reduced scatter of 0.54 dex.

This confirms recent results that the shape of the $\mcmo - \sigma$ relation is different for NSCs and 
super-massive black holes (SMBHs). We discuss this result in the context of the generalized idea of 
"central massive objects" (CMOs).

In order to assess which physical parameters drive the observed nuclear cluster masses, 
we also carry out a joint multi-variate power-law fit to the data. In this, we allow the nuclear cluster mass 
to depend on spheroid mass and radius (and hence implicitly on velocity dispersion), as well as on 
the size of the globular cluster reservoir. When considered together, the dependences on $\msph$ 
and $\rsph$ are roughly consistent with the virial theorem, and therefore $\mnsc \propto \sigma^2$.
However, the only statistically 
significant correlation appears to be a simple linear scaling between NSC mass and luminous spheroid mass.

We proceed to directly compare the derived NSC masses
with predictions for two popular models for NSC formation, namely i) globular cluster infall 
due to dynamical friction, and ii) in-situ formation during the early phases of galaxy formation
that is regulated via momentum feedback from stellar winds and/or supernovae. Neither model
can directly predict the observations, and we discuss possible interpretations of our findings.

\end{abstract}

\begin{keywords}
galaxies: nuclei -- galaxies: elliptical and lenticular, cD -- galaxies: photometry -- methods: statistical
\end{keywords}

\section{Introduction} \label{intro}

The presence of a central massive object in the nuclei
of galaxies appears to be a generic by-product of their formation
and evolution. Over the last decade, it has become clear that
in galaxies with massive stellar spheroids (i.e. massive ellipticals
or early-type spirals with bulges more massive than $\gtrsim 10^{10}\msun$), 
the nucleus is nearly always occupied by a super-massive black hole 
that outweighs by far the stellar mass within the central few parsecs.  
On the other hand, in most galaxies {\it without} massive stellar spheroids, 
such as disk-dominated spirals and low-mass ellipticals, an SMBH is only rarely 
present \citep{satyapal09}. Instead, the central mass concentration is dominated 
by a compact and massive nuclear star cluster. In between the two regimes, 
there appears to be a ``transitional'' spheroid mass range ($8\lesssim\,log(\msph[\msun]) \lesssim\,10$) 
over which the two CMO types can co-exist, and have comparable masses 
\citep{graham09,neumayer12}.

It is still unclear, however, whether these two incarnations of CMOs form 
through different physics, or are merely different evolutionary stages of a 
common formation mechanism. This question has triggered a very active area
of astrophysical research, with many observational studies of the
statistics and properties of NSCs \citep{carollo98,boeker02,boeker04,balcells03,grahamguzman03,walcher05,walcher06,cote06} 
and SMBHs \citep{satyapal08,gallo10} in low-mass spheroids, their co-existence \citep{grahamdriver07,seth08,gonzalez-delgado08},
as well as a number of proposed theories about the growth of NSCs \citep{capuzzo08,agarwal11,hartmann2011,antonini12} 
and the mutual feedback between NSC and SMBH \citep{mclaughlin06,nayakshin09}.

In 2006, three independent studies \citep{wehner06,rossa06,ferrarese06a} pointed
out that NSC masses appear to correlate with the masses of their host spheroids, confirming
an earlier result by \cite{grahamguzman03}. These scaling relations are similar to those
observed for SMBHs  \citep{ferrarese00,gebhardt00,haering04},
i.e. $\mcmo \propto \sigma^{\beta}$.
However, because of the observational difficulties to obtain reliable NSC masses, and
the relatively small range of spheroid masses over which NSCs dominate over SMBHs,
it is still debated whether both types of CMOs follow the same scaling relation (i.e. have the
same value of $\beta$), or whether they depend in different ways on the properties of their host spheroid.

While \cite{ferrarese06a} suggested that NSCs smoothly extend the SMBH relations
to lower masses, with $\beta \approx 4$ in both cases, a recent study by 
\cite{graham12} concluded that the two types of 
CMOs depend differently on the {\it total} spheroid mass: NSCs follow a relation with 
$\beta = 1.57\pm 0.24$ while for SMBHs, $4 \lesssim \beta \lesssim 5$. 

In order to add to the current discussion, we re-examine photometry of the stellar
spheroids and NSCs in 51 early-type galaxies, obtained with the {\it Hubble Space
Telescope} ACS Virgo Cluster Survey, and published in \cite{ferrarese06b} and
\cite{cote06}. The analysis of \cite{ferrarese06a} compared the \textit{luminous} mass 
of 29 NSCs to the {\it total} mass of their host galaxies, the latter being estimated 
from stellar velocity dispersion measurements. Here, we follow a somewhat different 
approach by comparing the {\it luminous} mass of \textit{both} NSC and galaxy spheroids, derived 
via color-corrected mass-to-light ratios. This not only yields more reliable stellar masses, but
also improves the statistics by increasing our sample size to 51 since we can use nearly the entire 
sample of NSCs identified in \cite{cote06}. 

This paper is structured as follows. In \S~\ref{sec:data}, we summarize the
data sources for our analysis and describe our method to convert the 
published photometry into estimates of the stellar
mass of both NSCs and host spheroids.  We present a quantitative analysis
of the resulting $\mnsc -\msph$ relation in \S~\ref{sec:corr}, as
well as of the inferred $\mnsc - \sigma$ relation. In addition, we
demonstrate through a multi-variate analysis that the only statistically significant 
correlation with any observed parameter is a simple linear scaling 
of NSC mass with spheroid mass.  
In \S~\ref{sec:predict}, we briefly summarize the theoretical predictions for
two models of NSC formation, compare the present-day NSC masses to those
predictions, and discuss the implications for the plausibility of either model.  
We conclude with a discussion and summary of our results and the implications 
for the origin of NSCs in \S~\ref{discussion} and \S~\ref{summary}, respectively. 

\section{The Data} \label{sec:data}
For our analysis, we use data for 51 nucleated early-type galaxies 
observed during the 
{\it Advanced Camera for Surveys} Virgo Cluster Survey \citep[ACSVCS,][]{cote04}. 
Taking advantage of the high spatial resolution of the {\it Hubble Space Telescope}, 
this data set provides detailed and accurate photometry for both the galaxy bodies 
and their NSCs. We reject five galaxies from the original sample of \citet{cote06} in
which the apparent NSCs are significantly offset from the galaxy's photocentre and
therefore, as discussed by \citet{cote06}, may well be globular clusters that only 
appear to reside close to the nucleus due to a chance projection.  

We note here that, given the morphological types of the sample 
galaxies (E, S0, dE, dS0, and dE,N), their stellar spheroids can be expected to 
be virialized, and to have little or no current star formation activity.
This justifies use of a single (color-dependent) mass-to-light ratio for
each galaxy spheroid in order to derive its stellar mass. 

We make use of the apparent z-band magnitudes, (g-z) colors, and half-light 
radii for both NSCs \citep[from][]{cote06} and host spheroids \citep[from][]{ferrarese06b}. 
In order to convert to absolute magnitudes and physical radii, we follow the
approach of \cite{cote06} and assume a common distance of 16.5\,Mpc for all galaxies.
For convenience, the observed galaxy and NSC magnitudes are listed again in Columns 2 and 
3 of Table~\ref{table:properties}.  We also list in Column~4 the (completeness-corrected) 
total number of globular clusters (GCs) as provided in \citet[][their Table 2]{peng08}
which we will use in \S~\ref{subsec:infall}.

In order to obtain estimates for the stellar masses of both NSCs and
spheroids, we need to multiply their respective z-band luminosities by an appropriate
mass-to-light ratio.  We use the empirically calibrated mass-to-light
ratios provided by \cite{bell03}, accounting for the (g-z) color of NSC
and spheroid, respectively, and list the resulting values for $\msph$ and
$\mnsc$ in Columns 5 and 8 of Table~\ref{table:properties}.  We also provide 
estimates for the average GC mass $\bar{m}_{\rm GC}$ for our sample of galaxies.  
This is done by multiplying the z-band galaxy masses 
by the percentage of the total mass of the galaxy in GCs (obtained
in the z-band) taken from \citet{peng08}.  This gives an estimate for
the total mass of the GC system, which we then divide by the total
number of GCs in order to arrive at an estimate for the average
GC mass.  Note that some of the percentages provided in \citet{peng08} 
are negative.  In these few cases, we leave blank the corresponding entry 
for $\bar{m}_{\rm GC}$ in Table~\ref{table:properties}.

\begin{table*}
\caption{Properties of the sample galaxies, identified by their VCS number (Column 1). 
Columns 2 and 3 list the absolute z-band magnitudes of the galaxy \citep{ferrarese06b} and NSC \citep{cote06} 
(after accounting for a distance of 16.5\,Mpc to get the absolute magnitude), respectively.  
Column 4 provides the completeness-corrected number of GCs ($N_{\rm GC}$) \citep{peng08}.  
Column 5 provides the total stellar mass of the galaxy $\msph$ calculated from its z-band 
magnitude; in units of $10^9\msun$.  Column 6 gives the effective radius $\rsph$ of the 
galaxy spheroid \citep{ferrarese06b}.  Column 7 lists the calculated velocity dispersion within 
the effective radius of the galaxy in $\kms$.  Column 8 gives the inferred stellar mass of the NSC 
$\mnsc$, column 9 its half-light radius \citep{cote06}, and column 10 its calculated velocity dispersion.  
Finally, Column 11 gives the average globular cluster mass $\bar{m}_{\rm GC}$ in units of $10^5\msun$.  
All listed errors denote the $1\sigma$ uncertainties.}
\begin{tabular}{|c|c|c|c|c|c|c|c|c|c|c|c|}
\hline
VCS  &     $M_z$    &   $M_z$   &        $N_{\rm GC}$       &        $\msph$       &  $\rsph$  &  $\sigma$  &        $\mnsc$       &  $R_{\rm NSC}$  &     $\sigma_{\rm NC}$     &    $\bar{m}_{\rm GC}$      \\
     &  (Galaxy)  &  (NSC)  &                             &  (10$^9$ $\msun$)  &    (kpc)    &   ($\kms$)  &  (10$^5$ $\msun$)  &    (pc)     &    ($\kms$)      &  (10$^5$ $\msun$)   \\
(1)  &     (2)    &   (3)   &         (4)       &          (5)         &          (6)           &     (7)     &        (8)       &           (9)          &    (10)     &           (11)         \\       
\hline  
 27  &  -20.68  &  -14.27  &   62.0  $\pm$  13.0  &  23.59  &  2.18  &  241.19  &   708.86  $\pm$  53.55  &   6.81  &  236.61 $\pm$  0.81  &  4.57  \\
 29  &  -20.19  &  -15.51  &   84.0  $\pm$  19.0  &  16.08  &  0.80  &  328.87  &  2188.37  $\pm$ 165.32  &  25.87  &  213.26 $\pm$  1.03  &  2.68  \\
 30  &  -20.73  &  -13.49  &   83.0  $\pm$  25.0  &  21.78  &  2.00  &  241.74  &   251.25  $\pm$  18.98  &   1.92  &  265.09 $\pm$ 28.65  &  4.20  \\
 31  &  -20.29  &  -12.95  &  116.0  $\pm$  24.0  &  17.83  &  1.34  &  267.95  &   231.43  $\pm$  17.48  &   2.80  &  210.68 $\pm$ 13.11  &  4.76  \\
 32  &  -20.26  &  -14.41  &   50.0  $\pm$  14.0  &  16.81  &  1.09  &  287.28  &   862.06  $\pm$  65.12  &  12.89  &  189.59 $\pm$  3.04  &  2.35  \\
 36  &  -20.0   &  -14.13  &   69.5  $\pm$   9.8  &  11.21  &  0.84  &  267.70  &   623.10  $\pm$  47.07  &  16.66  &  141.81 $\pm$  2.97  &   --   \\
 37  &  -20.10  &  -12.90  &   20.1  $\pm$   7.3  &   9.88  &  0.94  &  237.25  &   197.75  $\pm$  14.94  &   2.08  &  225.96 $\pm$ 21.88  &  2.95  \\
 38  &  -20.06  &  -15.37  &   47.0  $\pm$  11.0  &  13.93  &  1.05  &  267.09  &  2118.26  $\pm$ 160.02  &  40.13  &  168.47 $\pm$  1.03  &  3.56  \\
 39  &  -19.64  &  -17.14  &   72.0  $\pm$  12.0  &   8.45  &  1.29  &  188.07  &  8197.84  $\pm$ 673.67  &  62.47  &  277.04 $\pm$  1.03  &  4.58  \\
 42  &  -19.74  &  -15.14  &   71.0  $\pm$  14.0  &   8.74  &  1.15  &  201.91  &  1615.18  $\pm$ 122.01  &  47.81  &  134.77 $\pm$  4.30  &  3.82  \\
 44  &  -19.44  &  -11.58  &   52.3  $\pm$   8.5  &   7.15  &  0.89  &  207.37  &    38.14  $\pm$   2.88  &   4.81  &   65.32 $\pm$  1.34  &  2.19  \\
 47  &  -19.70  &  -12.02  &   58.6  $\pm$   9.3  &   8.60  &  1.56  &  172.21  &    89.90  $\pm$   6.79  &   4.24  &  106.71 $\pm$  3.02  &  3.23  \\
 48  &  -19.21  &  -12.81  &   35.1  $\pm$   7.6  &   4.09  &  1.61  &  116.75  &   142.51  $\pm$  10.77  &   2.88  &  163.02 $\pm$  9.69  &  3.38  \\
 49  &  -19.71  &  -12.48  &  114.0  $\pm$  12.0  &   6.94  &  1.22  &  175.08  &   113.26  $\pm$   8.56  &   3.28  &  136.17 $\pm$  6.48  &  3.41  \\
 50  &  -18.73  &  -12.09  &   24.9  $\pm$   6.0  &   3.85  &  1.76  &  108.50  &    73.43  $\pm$   5.55  &   2.80  &  118.67 $\pm$  7.38  &  3.25  \\
 51  &  -18.57  &  -10.79  &   17.2  $\pm$   5.4  &   2.77  &  1.01  &  121.20  &    21.05  $\pm$   1.59  &   2.48  &   67.52 $\pm$  5.07  &  3.06  \\
 52  &  -18.32  &  -13.61  &   10.4  $\pm$   5.0  &   2.94  &  0.55  &  168.80  &   304.45  $\pm$  23.00  &   8.65  &  137.56 $\pm$  0.74  &  5.93  \\
 55  &  -18.33  &  -12.46  &   48.7  $\pm$   8.4  &   2.67  &  0.96  &  122.24  &   100.97  $\pm$   7.63  &   3.04  &  133.56 $\pm$  7.26  &  5.98  \\
 57  &  -18.25  &  -13.24  &   43.4  $\pm$   7.9  &   2.00  &  1.34  &   89.75  &   196.63  $\pm$  14.85  &  12.25  &   92.88 $\pm$  1.38  &  2.54  \\
 58  &  -18.27  &   -9.90  &   21.3  $\pm$   6.1  &   2.10  &  0.74  &  123.27  &     7.76  $\pm$   0.59  &   1.76  &   48.67 $\pm$  5.90  &  1.77  \\
 59  &  -18.16  &  -11.02  &   10.8  $\pm$   5.6  &   1.77  &  2.44  &   62.41  &    23.81  $\pm$   1.80  &   3.04  &   64.85 $\pm$  3.52  &  0.82  \\
 60  &  -18.64  &  -12.20  &   66.0  $\pm$   9.5  &   3.05  &  1.63  &  100.27  &    88.16  $\pm$   6.66  &   2.16  &  148.05 $\pm$ 13.60  &  4.02  \\
 62  &  -18.34  &  -12.03  &   37.6  $\pm$   7.4  &   2.54  &  1.46  &   96.74  &    60.80  $\pm$   4.59  &   9.53  &   58.56 $\pm$  0.49  &  4.67  \\
 63  &  -18.19  &   -9.89  &   18.1  $\pm$   5.5  &   2.19  &  1.46  &   89.64  &    10.74  $\pm$   0.81  &  15.70  &   19.18 $\pm$  0.38  &  1.81  \\
 64  &  -18.25  &  -12.55  &   60.6  $\pm$   9.3  &   2.43  &  0.79  &  128.46  &   104.93  $\pm$   7.93  &  18.66  &   54.98 $\pm$  1.25  &  3.13  \\
 65  &  -18.04  &   -9.78  &   40.7  $\pm$   7.6  &   1.84  &  0.79  &  111.76  &     7.21  $\pm$   0.54  &   1.44  &   51.87 $\pm$  8.13  &  2.98  \\
 66  &  -18.32  &   -9.86  &   14.4  $\pm$   5.7  &   2.05  &  2.17  &   71.29  &    10.60  $\pm$   0.80  &   1.92  &   54.46 $\pm$  5.88  &  1.28  \\
 68  &  -18.69  &  -12.08  &   42.1  $\pm$   7.9  &   2.96  &  2.08  &   87.38  &    61.80  $\pm$   4.67  &   6.65  &   70.70 $\pm$  0.31  &  3.23  \\
 69  &  -18.18  &  -11.89  &   23.9  $\pm$   6.1  &   1.79  &  1.53  &   79.34  &    56.71  $\pm$   4.28  &   2.32  &  114.57 $\pm$  9.50  &  5.38  \\
 70  &  -17.01  &   -9.82  &    2.2  $\pm$   4.2  &   0.72  &  0.73  &   72.46  &     7.21  $\pm$   0.54  &   2.56  &   38.90 $\pm$  2.79  &   --   \\
 71  &  -17.80  &   -9.15  &   15.5  $\pm$   5.8  &   1.10  &  1.05  &   75.13  &     3.92  $\pm$   0.30  &   3.04  &   26.31 $\pm$  1.43  &  0.36  \\
 74  &  -17.34  &   -8.10  &    7.3  $\pm$   4.2  &   0.70  &  0.84  &   66.84  &     1.28  $\pm$   0.10  &   2.00  &   18.57 $\pm$  1.90  &  0.29  \\
 76  &  -17.29  &   -8.48  &    6.3  $\pm$   4.3  &   0.87  &  0.84  &   74.74  &     2.05  $\pm$   0.16  &   1.68  &   25.62 $\pm$  3.30  &  0.96  \\
 78  &  -17.74  &  -10.21  &   54.2  $\pm$   8.8  &   1.33  &  0.86  &   91.24  &    11.46  $\pm$   0.87  &   2.96  &   45.59 $\pm$  2.59  &  2.51  \\
 79  &  -18.14  &  -13.19  &   -6.0  $\pm$  13.0  &   2.35  &  0.57  &  149.06  &   197.79  $\pm$  14.94  &   9.69  &  104.75 $\pm$  0.93  &   --  \\      
 81  &  -17.55  &  -10.98  &   16.5  $\pm$   5.2  &   1.11  &  1.43  &   64.56  &    21.78  $\pm$   1.65  &   3.12  &   61.23 $\pm$  3.18  &  2.21  \\
 83  &  -17.39  &  -13.70  &    3.6  $\pm$   3.7  &   1.19  &  0.31  &  143.46  &   380.80  $\pm$  28.77  &  15.78  &  113.91 $\pm$  2.28  & 39.88  \\
 84  &  -17.66  &  -12.71  &   26.7  $\pm$   6.8  &   1.31  &  0.62  &  107.05  &   139.98  $\pm$  10.57  &   4.49  &  129.54 $\pm$  3.20  &  1.43  \\
 85  &  -16.96  &  -11.87  &   28.7  $\pm$   6.7  &   0.52  &  0.73  &   61.78  &    54.45  $\pm$   4.11  &   2.64  &  105.24 $\pm$  7.19  &  9.82  \\
 86  &  -17.15  &   -9.68  &    9.2  $\pm$   4.3  &   0.78  &  0.94  &   66.97  &     6.63  $\pm$   0.50  &   5.29  &   25.96 $\pm$  0.40  &  1.79  \\
 88  &  -17.92  &  -10.92  &    9.2  $\pm$   4.3  &   1.79  &  1.04  &   96.26  &    22.04  $\pm$   1.66  &   2.80  &   65.01 $\pm$  4.05  &  2.53  \\
 89  &  -17.47  &  -10.59  &   20.4  $\pm$   5.8  &   0.97  &  1.32  &   62.89  &    15.67  $\pm$   1.18  &   4.57  &   42.96 $\pm$  1.02  &  4.52  \\
 90  &  -17.10  &  -10.94  &    0.6  $\pm$   3.5  &   0.36  &  0.39  &   70.05  &    23.82  $\pm$   1.80  &   2.40  &   73.00 $\pm$  5.76  &   --   \\
 91  &  -17.43  &  -11.69  &   49.7  $\pm$   8.6  &   0.94  &  0.87  &   76.12  &    42.52  $\pm$   3.21  &  10.17  &   47.41 $\pm$  0.48  &  1.30  \\
 92  &  -16.93  &  -10.05  &    3.9  $\pm$   2.6  &   0.51  &  1.07  &   50.75  &     9.60  $\pm$   0.73  &   3.28  &   39.64 $\pm$  1.89  &  3.82  \\
 93  &  -16.94  &  -12.70  &   -9.0  $\pm$  14.0  &   0.80  &  0.27  &  127.22  &   142.87  $\pm$  10.79  &   5.05  &  123.38 $\pm$  2.19  &   --  \\   
 95  &  -17.12  &  -11.28  &   31.0  $\pm$   7.0  &   0.69  &  1.79  &   45.48  &    32.10  $\pm$   2.42  &  21.22  &   28.51 $\pm$  0.70  &  5.36  \\
 96  &  -17.37  &  -11.18  &   14.0  $\pm$   5.7  &   0.91  &  1.69  &   53.86  &    26.00  $\pm$   1.96  &   4.00  &   59.08 $\pm$  1.90  &  1.31  \\
 97  &  -16.71  &  -12.18  &    4.5  $\pm$   3.9  &   0.50  &  0.55  &   69.86  &    78.02  $\pm$   5.89  &   2.00  &  144.74 $\pm$ 14.80  &  0.89  \\
 99  &  -16.41  &   -9.58  &   11.7  $\pm$   4.8  &   0.34  &  0.89  &   45.22  &     5.61  $\pm$   0.42  &   4.65  &   25.49 $\pm$  0.58  &  1.33  \\
100  &  -17.40  &  -11.84  &   10.2  $\pm$   4.7  &   1.08  &  3.15  &   42.96  &    51.42  $\pm$   3.88  &   6.57  &   64.88 $\pm$  0.32  &  0.85  \\
\hline
\end{tabular}
\label{table:properties}
\end{table*}

\section{$\mnsc $ Scaling Relations} \label{sec:corr}
In this section, we present a quantitative analysis of the observed 
$\mnsc-\msph$ relation using estimates for \textit{luminous} masses 
derived from photometry.  We also consider the inferred $\mnsc-\sigma$ 
relation found by assuming virial equilibrium, in addition to a 
multi-variate analysis that quantifies the dependence of $\mnsc$ on 
the individual galaxy parameters.  In all cases, the fits are normalized 
by dividing the parameters being compared by their sample-averaged values.  
This significantly reduces the uncertainties in the offsets (i.e. the y-intercepts 
for our lines of best-fit).  All uncertainties given in this paper correspond to 
the $1\sigma$-confidence level.

\subsection{The $\mnsc - \msph$ correlation} \label{sec:mmcorr}
\begin{figure}
\begin{center}
\includegraphics[width=\columnwidth]{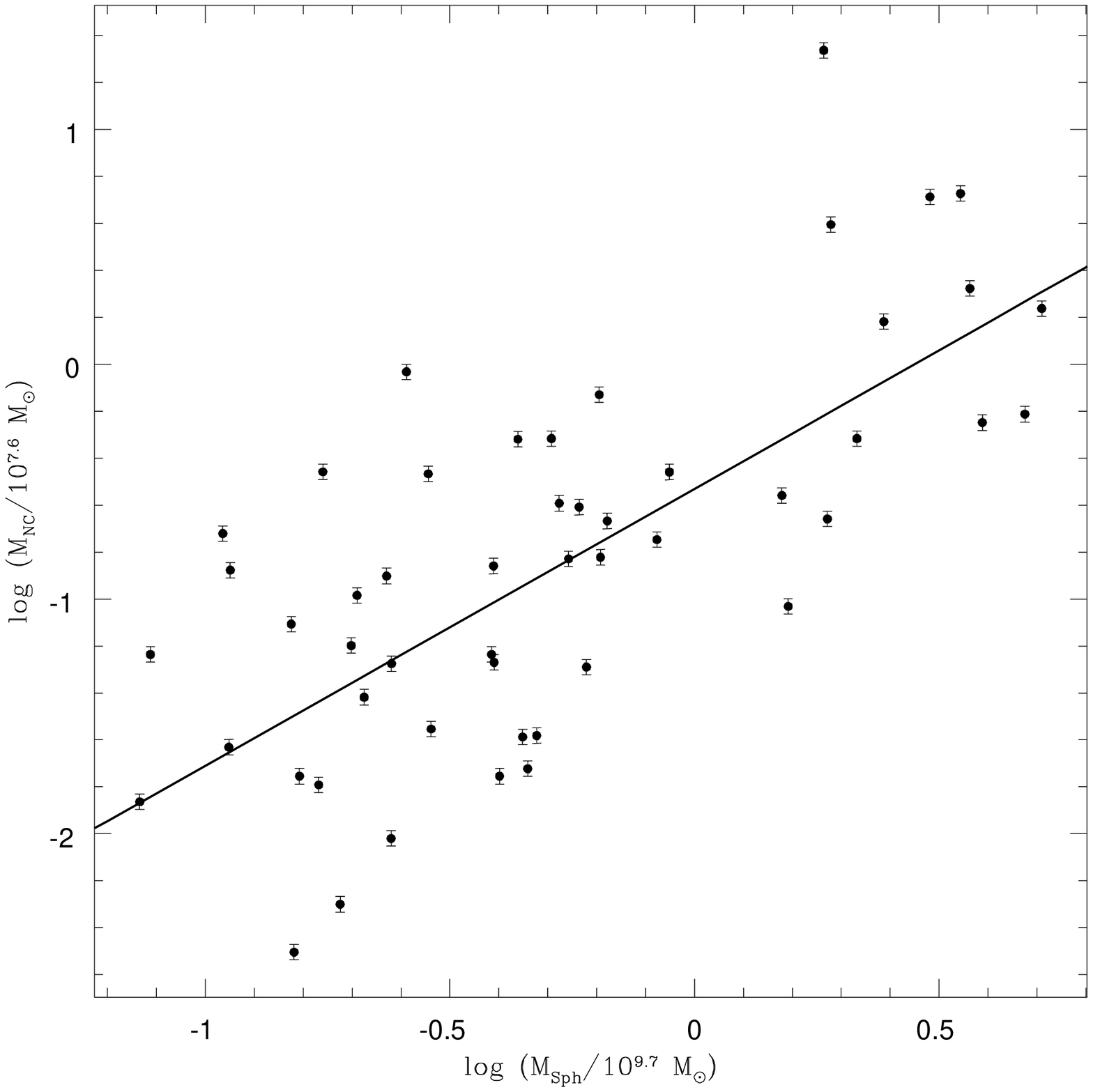}
\end{center}
\caption{Correlation between the stellar mass of NSCs and their host spheroids. The
solid line indicates the best-fitting linear relationship of Equation~\ref{eq:mmfit}.  
Error bars were calculated using the 0.041 mag uncertainty quoted by \citet{cote06}.}
\label{fig:mmcorr}
\end{figure}

Figure~\ref{fig:mmcorr} plots the stellar mass of the NSCs against that of their host spheroids. 
The solid line shows a weighted least-squares fit to the data found using a relation of the form
\begin{equation}\label{eq:mmfit}
log({\mnsc/10^{7.6} \msun}) = a + b \cdot log(\msph/10^{9.7} \msun).
\end{equation}
The uncertainties were calculated via a bootstrap methodology \citep{leigh11} 
in which we generated 1,000 fake data sets by randomly sampling (with replacement) 
data points from the observations.  We obtained lines of best fit for each 
fake data set, fit a Gaussian to the subsequent distribution and extracted its 
standard deviation.  
We find $a=-0.53 \pm 0.09$ and $b=1.18 \pm 0.16$, with a scatter of 0.57 dex.  A strong 
correlation between the two quantities is evident, with a Spearman correlation coefficient of 
$r_s=0.70$, a significance level at which the null hypothesis of zero correlation is 
disproved of $p_s=7.17\cdot 10^{-9}$, and a slope that is inconsistent with zero at the 
$7\sigma$ confidence level.  Table~\ref{table:properties} does not list uncertainties 
for the spheroid masses provided in column 5, since we did not have uncertainties for the 
z-band magnitudes from which they were calculated.  However, we 
performed a doubly-weighted least-squares fit using the spheroid masses and 
uncertainties provided in column 8 of Table 1 in \citet{peng08} (who derived their masses 
using different assumptions for the mass-to-light ratios than adopted in this paper).  The 
slope and y-intercept for the resulting line of best-fit are the same as found using 
the spheroid masses calculated in this paper to within one standard deviation.

This result is in broad agreement with various other studies. For example, using the $HST$ imaging data 
of \cite{lotz04} for 45 nucleated dE,N galaxies in Leo, Virgo and Fornax, \cite{wehner06} find values of 
$a=-3.10\pm 0.81$ and $b=1.18\pm 0.10$. Using $HST$ near-infrared images of early-type spiral bulges,
\cite{balcells07} find $a=-2.75\pm 0.15$ and $b=0.76\pm 0.13$. Our analysis, which should be an improvement 
over these earlier studies due to the larger sample size and the use of color-dependent mass-to-light ratios, 
therefore re-affirms an approximately linear dependence of $\mnsc$ on the {\it luminous} mass of the host 
spheroid.  

\subsection{The $\mnsc - \sigma$ correlation} \label{sec:mscorr}
As mentioned earlier, we assume that the sample galaxies are in 
dynamical equilibrium, and that, therefore, the virial theorem can be used to predict the
stellar velocity dispersion of the spheroid. We use the empirically calibrated formula 
of \cite{cappellari06}:
\begin{equation}\label{eqn:virial}
\sigma^2 = \frac{G\msph }{5f_g\rsph }
\end{equation}
where $\msph$ is the luminous mass of the galaxy spheroid, $\rsph$ is its half-mass radius,
and $f_g = \Omega_b/\Omega_m$ is the baryonic mass fraction. We assume that $f_g=0.16$ 
\citep{spergel03} for all sample galaxies, and that the measured half-light radii (Column\,6 of 
Table~\ref{table:properties}) are a good approximation of the true half-mass radii, i.e. that 
luminous and dark matter follow the same radial distribution.
\begin{figure}
\begin{center}
\includegraphics[width=\columnwidth]{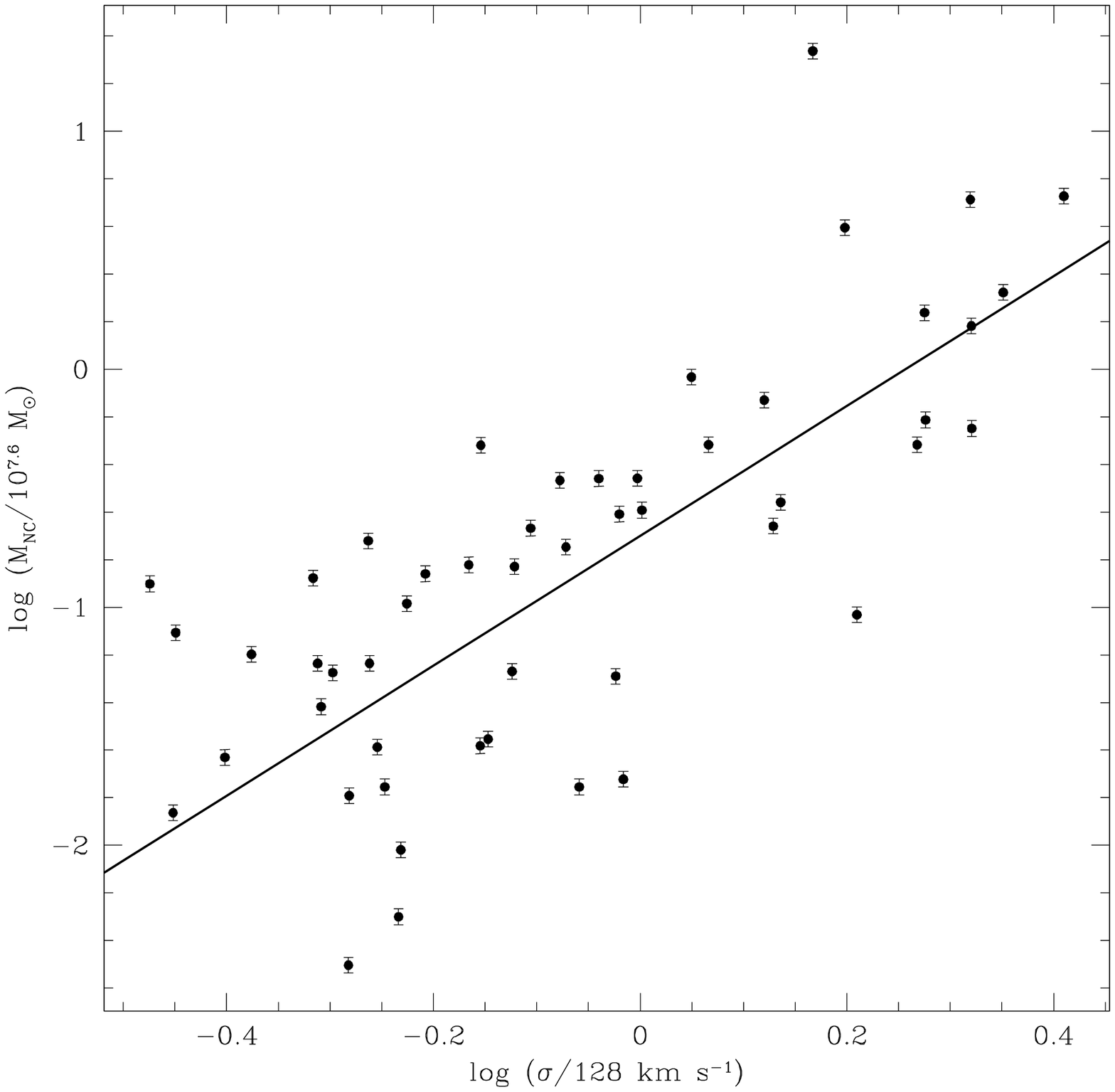}
\end{center}
\caption{Correlation between the stellar mass of NSCs and the {\it predicted}
velocity dispersion of their host spheroids. The solid line indicates the best-fitting 
linear relationship of Equation~\ref{eq:msfit}.  Error bars are described as in 
Figure~\ref{fig:mmcorr}.}
\label{fig:mscorr}
\end{figure}
The resulting values for the predicted velocity dispersion are listed in Column\,7 
of Table~\ref{table:properties}. 
Figure~\ref{fig:mscorr} then plots the relation between $\mnsc$ and $\sigma$.
The best fit is again indicated by the solid line, it is described by
\begin{equation}\label{eq:msfit}
log\big( \frac{\mnsc}{10^{7.6} \msun}\big) = (-0.70\pm 0.07) + (2.73\pm 0.29) \cdot log\big( \frac{\sigma}{128\kms}\big) .
\end{equation}
This correlation is slightly stronger ($r_s=0.72$ and $p_s=1.48\cdot 10^{-9}$) than that 
between $\msph$ and $\mnsc$, and has a slightly reduced scatter (0.54 dex).

The best-fit slope disagrees with that of \cite{ferrarese06a} who
conclude, based on {\it measured} velocity dispersions of a smaller subset (29 objects)
of the sample discussed here, that 
$log({\mnsc}) = (6.91\pm 0.11) + (4.27\pm 0.61) \cdot log(\sigma /54\kms)$.  
Taken at face value, this disagreement implies that our above assumptions
are wrong, and that either the half-light radius is a poor indicator of the half-mass 
radius, or that $f_g$ varies across the galaxy sample. For example, in order 
for Equation\,\ref{eqn:virial} to yield the slope of \cite{ferrarese06a}, $f_g$ would 
have to scale with the square root of the \textit{luminous} spheroid mass.

On the other hand, it is also possible that the relatively small sample used for
the \citet{ferrarese06a} analysis causes systematic errors in the derived slope
of the fit. This last possibility appears to be supported by a recent study 
of \citet{graham12} 
who re-analyzed the \citet{ferrarese06a} data, together with complementary 
literature data, and concluded that $\mnsc \propto \sigma^{1.57\pm 0.24}$, 
in better agreement with Figure~\ref{fig:mscorr}.

\subsection{Multi-variate Analysis}\label{subsec:multivar}
We now attempt to quantify whether or not the host galaxy properties are
correlated in a way that is consistent with our use of the virial theorem
as described in the previous section. More specifically, 
Equation\,\ref{eqn:virial} implies that the (total) mass of the spheroid and 
its half-mass radius are correlated. If the luminous matter indeed follows 
the same radial distribution as the dark matter, this should also be true for 
the \textit{luminous} spheroid mass and its \textit{half-light} radius $\rsph$. 

To assess whether or not this is indeed the case, we perform a multi-variate 
analysis for our sample, using a generalized relation of the form
\begin{equation}
\label{eqn:model_gen}
\mnsc \sim {\epsilon}\msph^{\alpha}\rsph^{\beta}N_{\rm GC}^{\delta},
\end{equation}
where $\alpha$, $\beta$, $\delta$, and $\epsilon$ are all free parameters.  

Note that we have included the total number of GCs, $N_{\rm GC}$, in this analysis, 
since the dependence of $\mnsc$ on this parameter will be tested in \S\,\ref{sec:predict}.  
We perform a direct weighted least-squares fit to 
Equation~\ref{eqn:model_gen}, and calculate 2$\sigma$ and 
3$\sigma$ joint confidence intervals for the correlations between 
possible pair of parameters $\alpha$, $\beta$, and $\delta$.  This is done 
by calculating the chi-squared value for every combination of each 
parameter pair, and identifying the minimum.  We then find every combination 
that gives a chi-squared value that is within the specified confidence 
interval of the minimum in order to obtain our joint confidence intervals.  
For a description of a similar application of this technique, we refer the 
reader to \citet{verbunt07}.

Our goal is to determine which of the parameters in Equation~\ref{eqn:model_gen} 
are correlated with $\mnsc$ at a statistically significant level while 
naturally accounting for any correlations between them.  
In this way, we are able to quantify the degree to which $\msph$ and $\rsph$ 
are driving the correlation between $\mnsc$ and $\sigma$.
\begin{figure}
\begin{center}
\includegraphics[width=\columnwidth]{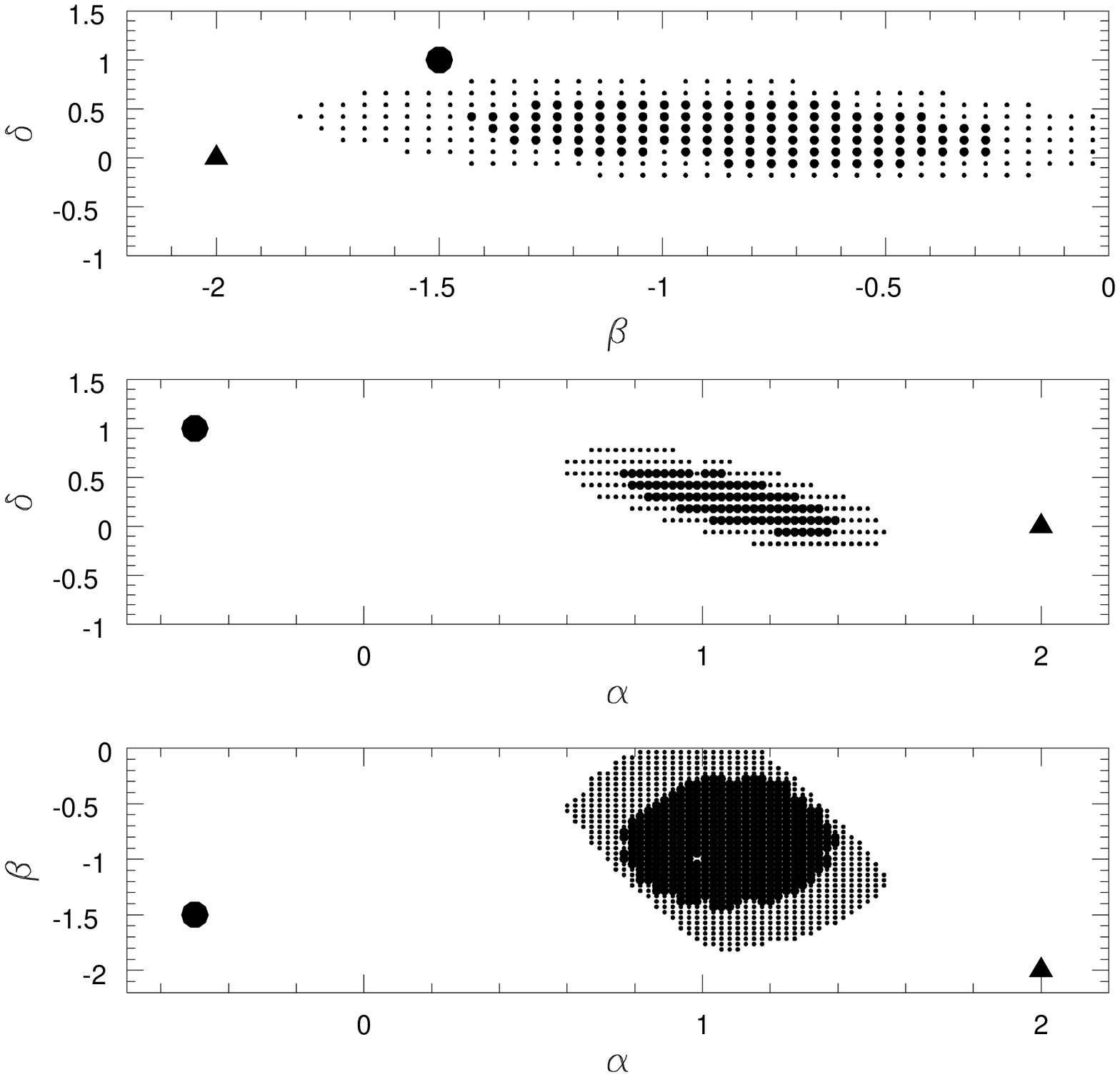}
\end{center}
\caption[Joint confidence intervals for three projections of the 
4-dimensional parameter space in Equation~\ref{eqn:model_gen}.]{The 
large and small filled circles show the 1$\sigma$ and 2$\sigma$, respectively,
joint confidence intervals on the two parameters being considered for all 
three projections of the 4-dimensional parameter 
space in Equation~\ref{eqn:model_gen}.  The large solid 
circles and triangles show the exact power-law predictions for the GC 
infall and the in situ star formation models, respectively.  
\label{fig:contours-nc}}
\end{figure}

The results are shown in Table~\ref{table:bestfit-nc} and illustrated in 
Figure\,\ref{fig:contours-nc}.  As expected given the results of 
\S\,\ref{sec:mmcorr}, we find a roughly linear dependence of $\mnsc$ on $\msph$.  
If the virial theorem is valid throughout our galaxy sample, 
$\mnsc \propto \msph$ implies that $\mnsc \propto \rsph^{-1}$.  
The results for $\rsph$ appear to be consistent with this notion, 
but we caution that the uncertainty for this parameter is large.  
Therefore, our results suggest that $\msph$ is the only galaxy 
parameter that correlates significantly with $\mnsc$.  

\begin{table}
\caption{Best-fitting power-law indices for the parameters of
Equation~\ref{eqn:model_gen}.}
\begin{tabular}{|c|c|}
\hline
Parameter          &    Best-Fitting Value   \\
\hline
log $\epsilon$     &   -1.0$^{+2.6}_{-3.0}$  \\
$\alpha$           &   1.1$^{+0.2}_{-0.3}$  \\
$\beta$            &   -0.9$^{+0.6}_{-0.5}$  \\
$\delta$           &   0.3$^{+0.3}_{-0.4}$  \\
\hline
\end{tabular}
\label{table:bestfit-nc}
\end{table}

\section{Comparison with Proposed Formation Scenarios} \label{sec:predict}
In this section, we compare theoretical predictions for NSC masses 
from two popular formation models to the observational data.
We begin by deriving 
theoretical relations for each formation scenario, which we use as a basis 
for comparison with observational data.  
\subsection{In-Situ Formation Regulated by Momentum Feedback} \label{subsec:feedback}
A plausible mechanism for the birth of ``seed'' NSCs, proposed by \citet{mclaughlin06}, 
is an extension of the arguments presented in \citet{king03} and \citet{king05} for 
the $\mbh$-$\sigma$ relation.  
It involves the formation of a massive star cluster from gas accumulating at the 
center during the early phases of the galaxy's evolution.
Very soon after the onset of star formation, stellar winds from massive 
stars and supernovae drive a superwind from the nucleus.  Initially, the 
wind is momentum-conserving with a momentum flux that is proportional 
to the Eddington luminosity.  It sweeps up the surrounding medium
into a thin supershell that gets pushed outward by the ram pressure it 
generates.  Eventually, gravity balances the ram pressure and the shell
stalls.  More gas then falls into and feeds the nucleus.  The
process is repeated until the NSC reaches a critical mass, at which point
the stall radius becomes very large (since the stall radius is proportional 
to the mass of the NSC).  This gives the gas time to cool
before it can reach the stall radius, since the cooling time begins to exceed the
crossing time of the bubble.  The gas becomes energy-conserving, and 
accelerates in its expansion to escape from the galaxy.  This fixes
the mass of the NSC at the critical mass \citep{mclaughlin06}:
\begin{equation}
\label{eqn:m_crit}
\mnsc = \frac{f_g{\kappa}{\sigma}^4}{{\lambda}{\pi}G^2},
\end{equation}
where $f_g = 0.16$ again is the baryonic mass fraction, 
$\kappa = 0.398$ cm$^2$ g$^{-1}$ is the electron
scattering opacity, and $\sigma$ is the velocity dispersion of the
spheroid. If correct, this mechanism would therefore naturally explain
the $\mnsc \propto \sigma^4$ dependence suggested by \cite{ferrarese06a}.

\citet{mclaughlin06} propose a value of $\lambda \sim 0.05$ 
for the efficiency of massive-star feedback from a young nuclear cluster.  
This is based on simple calculations for the contributions from both 
supernovae and stellar winds to the total momentum flux, and agrees 
with more detailed calculations of \citet{leitherer92} for the 
total momentum deposition from solar-metallicity stars. 

We can re-write Equation~\ref{eqn:m_crit} using the virial theorem.  
This assumes that the present-day mass distribution of the galaxy is representative 
of its distribution at the time of NSC formation. In particular, if 
the NSC formed before the galaxy mass was fully assembled, 
the velocity dispersion of the surrounding spheroid may well have been 
lower than it is today.  

Substituting the virial relation given by Equation~\ref{eqn:virial} into 
Equation~\ref{eqn:m_crit}, this predicts for the NSC mass:
\begin{equation}
\label{eqn:M_nc_1}
\mnsc = \frac{{\kappa}}{25f_g{\lambda}{\pi}}\msph^2\rsph^{-2}.
\end{equation}
Re-writing this in a more convenient form:
\begin{equation}
\label{eqn:M_nc_1b}
\mnsc = 1.66 \times 10^6 \msun \Big( \frac{1}{\lambda} \Big) \Big( \frac{2\kpc}{\rsph} \Big)^2 \Big( \frac{\msph}{10^9 \msun} \Big)^2,
\end{equation}
where we have assumed the baryonic mass fraction $f_g = 0.16$,
and the electron scattering opacity $\kappa = 0.398$ cm$^2$ g$^{-1}$.

\subsection{Infall and Merging of Globular Clusters} \label{subsec:infall}
The second NSC formation scenario occurs 
via the successive mergers of globular clusters (GCs) that spiral into
the galactic centre due to dynamical friction \citep[e.g.][]{tremaine75,
quinlan90}. Using N-body simulations of merging GCs in galaxy nuclei, 
several studies have successfully reproduced some of the properties
of NSCs, including their small effective radii and central 
velocity dispersions \citep[e.g.][]{capuzzo08}.  

In general, the time required for a GC of mass $m_{\rm GC}$ that starts on 
a circular orbit with radius $r$ from the galactic centre to spiral into the 
nucleus is given by \citep{binney87}:
\begin{eqnarray}
\label{eqn:tdyn}
\tdyn \, {\rm [yr]} &=& \frac{1.65}{G\ln{\Lambda}}\frac{r^2{\sigma}}{m_{\rm GC}} \\
           &=& \frac{3.73 \times 10^{11}}{\ln{\Lambda}}\Big( \frac{r}{2\kpc} \Big)^2 \Big( \frac{\sigma}{250 \kms} \Big) \Big( \frac{10^6 \msun}{m_{\rm GC}} \Big) \nonumber
\end{eqnarray}
where $\sigma$ is the velocity dispersion of the galaxy.  
The Coulomb logarithm is denoted by $\ln{\Lambda}$, and is 
on the order of 10 \citep{binney87}.  According to Equation~\ref{eqn:tdyn}, 
dynamical friction operates most efficiently on the most massive GCs.  

If the NSC was indeed built from GCs that have sunk into the nucleus, its mass
should follow a relation of the form:
\begin{equation}
\label{eqn:M_nc_2a}
\mnsc = f_{\rm GC}M_{\rm GC} \sim f_{\rm GC}N_{\rm GC}\bar{m}_{\rm GC},
\end{equation}
where $M_{\rm GC}$ is the total 
mass of the globular cluster system, $N_{\rm GC}$ is the total 
number of GCs, $\bar{m}_{\rm GC}$ is the {\it present-day} average 
GC mass of the galaxy (Column\,11 of Table\,\ref{table:properties}), 
and $f_{\rm GC}$ is the mass fraction of the GC system 
that has merged onto the NSC.  

This mass fraction should be proportional to the ratio of the age 
of the galaxy, $\tgal$, to the time for an average GC to fall 
into the core due to dynamical friction, $\tdyn$:
\begin{equation}
\label{eqn:M_nc_2b1}
\mnsc =f_{\rm tot}\frac{\tgal}{\tdyn}\, N_{\rm GC} \, \bar{m}_{\rm GC} .
\end{equation}
Here, the additional constant $f_{\rm tot}$ accounts for the fact that only GCs that
are located within a radius at which $\tdyn \leq \tgal$ will 
have had enough time to have fallen into the NSC via dynamical friction. 

Substituting $\tdyn = \tgal = 10\gyr$ in Equation\,\ref{eqn:tdyn} shows that
for most galaxies in our sample, only GCs 
well inside the half-light radius $\rsph$ can be expected to have fallen into the nucleus.  
However, the present-day orbits of most GCs in our sample are comparable 
to or exceed the half-light radius in their host galaxies \citep{peng08}. 
In other words, most GCs will not have had sufficient time to have spiraled 
into the nucleus, i.e. $f_{\rm GC} \ll 1$. The compilation of \cite{peng08} 
can be used to estimate the present-day value of $f_{\rm tot}$
by calculating the fraction of GCs inside $\rsph$. This yields $f_{\rm tot} \approx 0.2$,
in reasonable agreement with the notion that $N_{\rm GC}$ has not changed 
significantly over the lifetime of the galaxy.

We assume a universal initial GC MF that adheres to the same
radial dependence in all galaxies, and use the dynamical friction 
timescale calculated at the half-light radius as a proxy for the 
corresponding volume-integrated value.  

Using Equation~\ref{eqn:tdyn}, we get from Equation~\ref{eqn:M_nc_2b1}:
\begin{equation}
\label{eqn:M_nc_2b2}
\mnsc = \frac{G\ln{\Lambda}}{1.65}\tau_{\rm gal}\bar{m}_{GC}^2\rsph^{-2}\sigma^{-1}f_{\rm tot}N_{\rm GC},
\end{equation}
where we have used the half-light radius of the galaxy $\rsph$ as 
a proxy for the average initial orbital radius from which clusters 
begin spiraling into the nucleus.  

Using the virial theorem to substitute for $\sigma$ (Equation~\ref{eqn:virial}), 
and $f_{\rm tot} = 0.2$ then yields 
\begin{equation}
\label{eqn:M_nc_2b3}
\mnsc = 0.43(Gf_g)^{1/2}\ln{\Lambda}\tau_{\rm gal}\bar{m}_{\rm GC}^2\msph^{-1/2}\rsph^{-3/2}N_{\rm GC}.
\end{equation}
Alternatively, we can re-write Equation~\ref{eqn:M_nc_2b3} in the more convenient form:
\begin{equation}
\label{eqn:M_nc_2b4}
\mnsc = 5.36 \cdot 10^6\msun \Big( \frac{\rm 2\kpc}{\rsph} \Big)^2 \Big( \frac{250 \kms}{\sigma} \Big) \Big( \frac{\bar{m}_{\rm GC}}{10^6 \msun} \Big)^2 \Big( \frac{N_{\rm GC}}{10^2} \Big),
\end{equation}
where we have assumed a galaxy age of 10 Gyrs, $f_g = 0.16$, and 
$\ln{\Lambda} = 10$.

\subsection{Comparison with Observations} \label{subsec:comparison}
We proceed to compare the predictions of the two formation scenarios
discussed in the previous sections to the observed NSC masses, $\mnsc^{\rm obs}$.

\begin{figure}
\begin{center}
\includegraphics[width=\columnwidth]{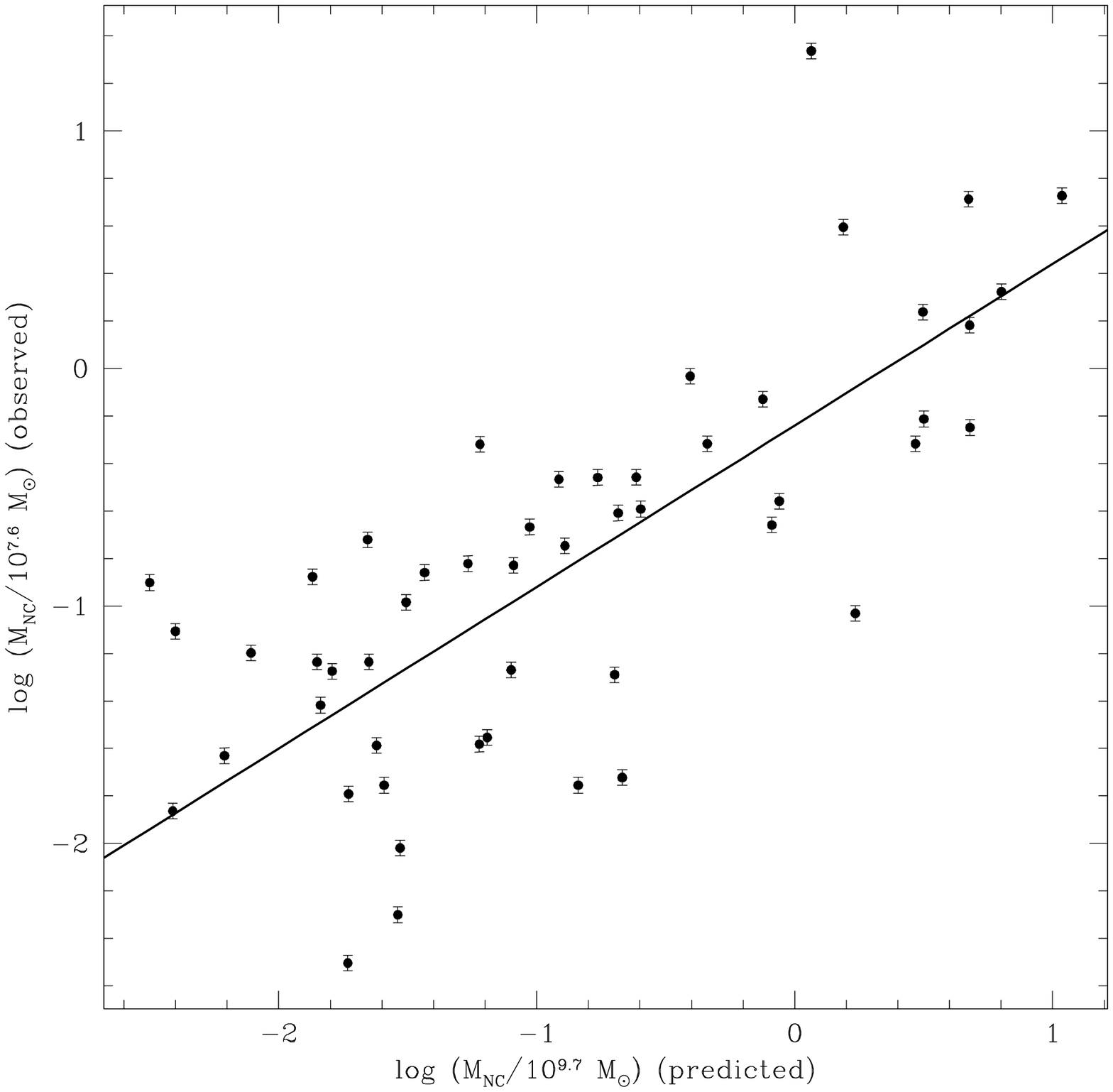}
\end{center}
\caption{Observed versus predicted NSC masses for the in situ star formation
model, with the latter calculated using Equation~\ref{eqn:M_nc_1}.  The solid line shows
a weighted least-squares fit to the data according to Equation\,\ref{eqn:wlsq_sf}.
Error bars are described as in Figure~\ref{fig:mmcorr}.
\label{fig:obs_vs_pred_sf}}
\end{figure}

For the in-situ star formation model, the predicted NSC masses, $\mnsc^{\rm pred}$, 
as calculated from Equation~\ref{eqn:M_nc_1}) are plotted against
the observations in Figure~\ref{fig:obs_vs_pred_sf}.
A weighted least-squares fit yields the relation
\begin{equation}
\label{eqn:wlsq_sf}
log \Big( \frac{\mnsc^{\rm obs}}{10^{7.6} \msun}\Big) = (0.68 \pm 0.07)\,log \Big( \frac{\mnsc^{\rm pred}}{10^{9.7} \msun}\Big) - (0.24 \pm 0.09), 
\end{equation}
The correlation is rather strong ($r_s =0.75$, $p_s = 2.34\cdot 10^{-10}$),
and has a scatter of 0.54 dex. 
However, the offset between the observed and predicted values 
suggests that the in-situ formation model \textit{over-predicts} the observations 
by roughly \textit{two orders of magnitude} across the galaxy sample. 
Given that secular evolution processes such as additional infall of 
cold molecular gas \citep[e.g.][]{sch06,sch07} and/or young star clusters
may well have added to the observed present-day NSC masses, and that
all mechanisms for mass loss, such as the ejection of stars due to dynamical 
encounters, appear negligible in NSCs \citep[e.g.][]{spitzer87, heggie03, valtonen06} 
this appears to present a severe challenge for the \cite{mclaughlin06}
model.

\begin{figure}
\begin{center}
\includegraphics[width=\columnwidth]{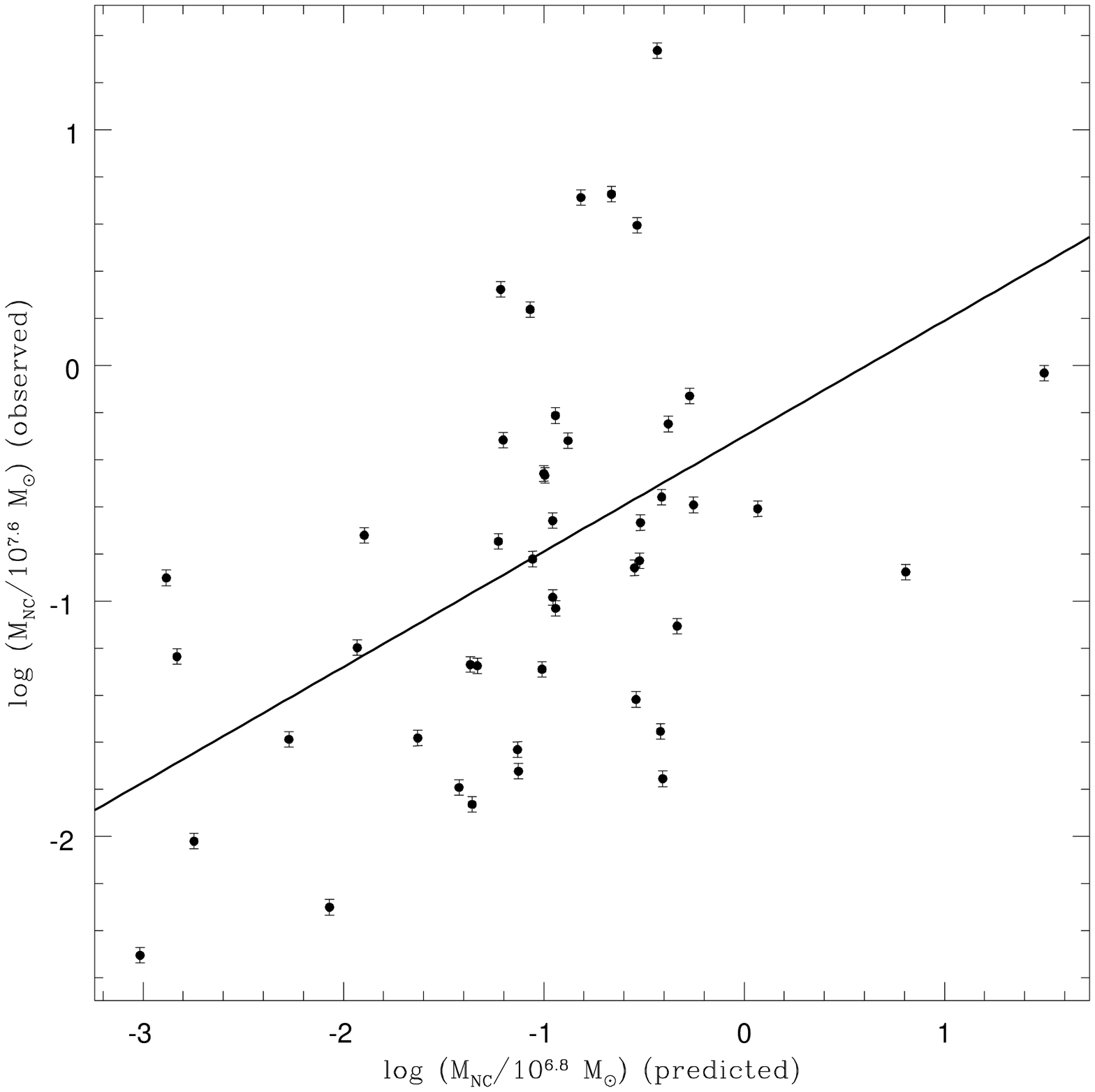}
\end{center}
\caption{Observed versus predicted NSC masses for the GC infall model, with
the latter calculated using Equation~\ref{eqn:M_nc_2b3}. 
The solid line shows a weighted least-squares fit to the data according to 
Equation\,\ref{eqn:wlsq_gc}.  Error bars are described as in 
Figure~\ref{fig:mmcorr}.
\label{fig:obs_vs_pred_gc}}
\end{figure}

Figure~\ref{fig:obs_vs_pred_gc} makes the corresponding comparison 
for the GC infall model. Here, the predicted NSC masses were calculated 
using Equation~\ref{eqn:M_nc_2b3}. 
The solid line in Figure~\ref{fig:obs_vs_pred_gc} is the best fit linear relation
between the two quantities, described by
\begin{equation}
\label{eqn:wlsq_gc}
log \Big( \frac{\mnsc^{\rm obs}}{10^{7.6} \msun}\Big) = (0.42 \pm 0.26)\, log\Big( \frac{\mnsc^{\rm pred}}{10^{6.0} \msun}\Big) - (0.74 \pm 0.13).
\end{equation}
The best-fit correlation has a slope that is marginally consistent with zero, has a large
scatter of 0.80 dex, and is statistically weak ($r_s =0.49$, $p_s = 5.24\cdot 10^{-4}$).
Moreover, the y-intercept suggests that the GC infall model \textit{under-predicts} the
observed NSC masses by about an order of magnitude across the galaxy sample. 
Adding to this problem is the fact that, as pointed out by \cite{agarwal11}, many 
GCs may be disrupted before reaching the nucleus, which would result in an even
smaller NSC mass. We conclude that the GC infall model provides a poor description 
of the observations.

The multi-variate analysis presented in \S\,\ref{subsec:multivar} 
offers an additional test of the in situ star formation and GC 
infall models.  In particular, from Equation~\ref{eqn:M_nc_1}, the in situ star
formation model predicts $\alpha = 2$, $\beta = -2$, and $\delta = 0$, 
whereas, from Equation~\ref{eqn:M_nc_2b3}, the GC infall model predicts 
$\alpha = -1/2$, $\beta = -3/2$, and $\delta = 1$.  
Therefore, if the various combinations of these values for $\alpha$, $\beta$,
and $\delta$ for one of our competing formation scenarios are found
to lie within the joint confidence intervals in every 2-dimensional
projection of this 3-dimensional parameter space, this could be
interpreted as evidence that the data are consistent with that
particular formation scenario.

The 1$\sigma$ and 2$\sigma$ joint confidence intervals for all three 
projections of our 3-dimensional parameter space are shown in 
Figure~\ref{fig:contours-nc}.  The theoretically predicted 
relations are shown by solid triangles and circles for the in situ 
star formation and GC infall models, respectively, 
as described in the figure caption.  The precise values for the 
power-law indices corresponding to both the 
in situ and GC infall models do not coincide with even the 2$\sigma$ 
confidence intervals in any of the three projections. 

\section{Discussion} \label{discussion}
Our results suggest that $\mnsc$ scales linearly with luminous spheroid mass 
$\msph$.  We also find a dependence of the form $\mnsc \propto \sigma^{2.73\pm 0.29}$ 
when using velocity dispersions estimated via the virial theorem. 
This appears to support the conclusion of \citet{graham12} that the 
observed $\mnsc-\sigma$ relation is \textit{not} an extension 
of the observed $\mbh-\sigma$ relation. Our results thus add further evidence 
to the notion that NSCs and SMBHs may not share a common origin, and that, 
instead, different formation mechanisms may be responsible for the two incarnations
of a central massive object.  We note that we find a slightly stronger 
correlation for the $\sigma-\mnsc$ relation than we do for the $\msph-\mnsc$ 
relation. This raises the question whether perhaps a more fundamental
relation is that between the velocity dispersion of the host spheroid and that of
the NSC.

\begin{figure}
\begin{center}
\includegraphics[width=\columnwidth]{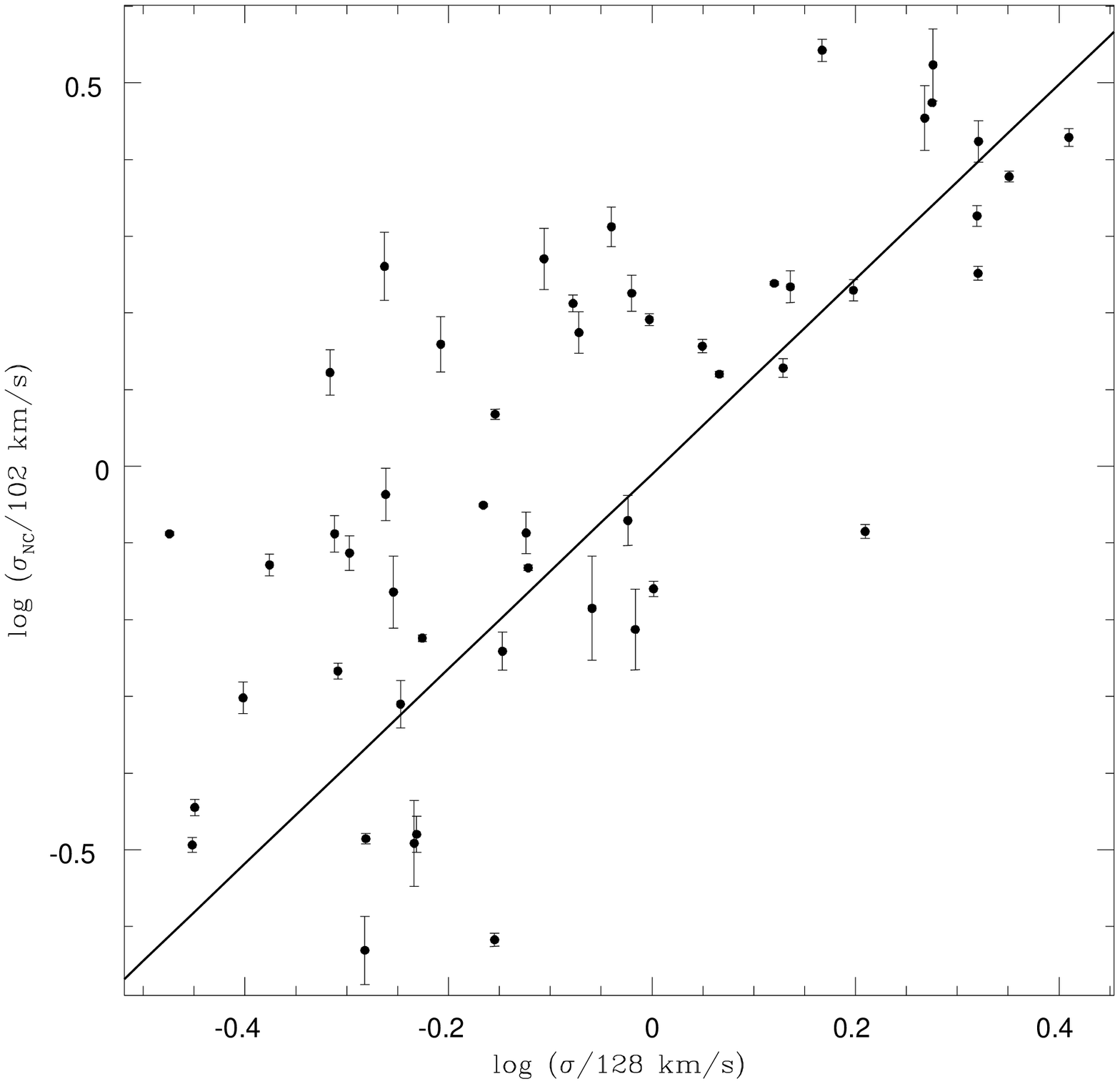}
\end{center}
\caption[]{
Comparison of the stellar velocity dispersion for NSC and host spheroid.  
The solid line shows a weighted least-squares fit to the data according to Equation\,\ref{eqn:sigsig}.  
Error bars were calculated using the 0.041 mag and 0.007 arcsecond uncertainties 
quoted by \citet{cote06} for the NSC apparent magnitudes and radii, respectively.
\label{fig:sigsig}}
\end{figure}

Such a relation could be explained if the NSCs are in thermal equilibrium with 
their host spheroids, possibly because galaxies tend to evolve towards (or be 
born in) such a state. To test this, we again used the virial theorem to calculate 
velocity dispersions also for all NSCs in the galaxy sample, using their effective 
radii measured by \cite{cote06}, and listed in Column\,9 of Table~\ref{table:properties}.
Figure\,\ref{fig:sigsig} shows that both quantities are indeed correlated,
with a linear fit yielding 
\begin{equation}
\label{eqn:sigsig}
log \Big( \frac{\sigma_{\rm NC}}{102\kms}\Big) = (1.27 \pm 0.09)\,log \Big( \frac{\sigma}{128\kms}\Big) - (0.01 \pm 0.02), 
\end{equation}
The correlation is rather strong ($r_s =0.73$), and has a smaller scatter (0.24\,dex) around the best linear fit
than either the  $\msph-\mnsc$ or the $\sigma-\mnsc$ relation.  
This supports the idea that the the dynamical evolution of NSCs is significantly affected by heat input 
from the surrounding galaxy \citep[e.g.][]{merritt09}. If indeed the present-day NSCs have reached an energy 
balance with their host spheroids, it implies that the time-scale for energy exchange between them is short 
compared to the age of the galaxy. This suggests that the present-day properties of NSCs do not necessarily 
reflect those at the time of their formation, since there was ample time for them to be modified by the 
presence of the surrounding spheroid.

The masses we have used for our analysis are derived from photometry, 
and therefore measure only the stellar mass of the galaxy. 
Measured velocity dispersions, on the other hand, provide a 
better proxy for the total mass of the spheroid, including the dark 
matter halo. It would therefore be most useful to perform a similar 
analysis using measured velocity dispersions once they become
available. 
Nevertheless, because of the arguments presented in Section~\ref{subsec:multivar}, 
our results have cast some doubt on the assumption that 
the baryonic mass fraction $f_g$ is the same for all early-type galaxies 
and/or that the dark matter has the same radial profile as the luminous 
matter. 

We have also compared the observed present-day NSC masses to theoretical 
predictions from two proposed models for NSC formation, namely in-situ star formation 
regulated by momentum feedback and the successive infall of star
clusters due to dynamical friction.  
As discussed in \S\,\ref{subsec:comparison}, both models appear to have
difficulties in predicting the observations.

The in-situ model predicts NSC masses that exceed the observed ones by about two orders of magnitude. 
Since the NSC may well have grown through secular processes throughout the life of the galaxy,
this is probably only a lower limit for the true difference. The discrepancy may be somewhat 
reduced if one assumes that the host galaxy was significantly less massive and/or more extended 
at the time of NSC formation than it is today. In this case, Equation~\ref{eqn:M_nc_1} would
yield lower NSC masses. For example, if the NSC host was
half as massive and twice as extended when the NSC formed, the resulting NSC mass should
be $\sim 16$ times lower. On the other hand, in this scenario it is difficult to
explain why the $\mnsc - \sigma$ relation does not seem to have the predicted
slope of $\beta = 4$ \citep{graham12}.

The GC infall model, on the other hand, systematically under-predicts the
observed NSC masses by about an order of magnitude across the galaxy sample. 
This could, at least partly, be rectified by adopting a larger value for the average GC mass.  
Indeed, the most massive GCs are most responsive to dynamical friction, and
therefore are most likely to have spiraled into the nucleus by today.
Using the present-day average GC mass as an estimate for the average mass of 
the GCs that have merged onto the NSC in Equation~\ref{eqn:M_nc_2b3} may
therefore provide too low an estimate. Indeed, most NSC masses can be reproduced
if ~10 very massive GCs ($\gtrsim 10^6 \msun$) had fallen in via dynamical friction.  
With the data at hand, we cannot rule out this possibility.

In summary, both models appear to have difficulties in explaining the observed
NSC masses. Of course, this does not imply that neither in-situ star formation 
nor GC infall has occurred, but it suggests that either model by itself provides
an over-simplified description of NSC formation.  

\section{Summary} \label{summary}
In this paper, we have performed a statistical analysis to study the
observed $\mnsc-\msph$ and inferred $\mnsc-\sigma$ relations
using a large sample of 51 nucleated early-type galaxies taken from
the ACSVCS.  We used \textit{luminous} galaxy
masses, and calculated velocity dispersions found by assuming virial
equilibrium applies for our sample of galaxies.  We confirm a linear 
relationship between NSC mass and luminous host spheroid mass reported
by previous studies, but infer a shallower slope of the $\mnsc - \sigma$
relation than reported in \cite{ferrarese06a}.

Our results support a recent claim \citep{graham12} that 
NSCs do not obey the same $M-\sigma$ relation as do SMBHs, 
adhering instead to a shallower dependence. Specifically, we found 
$\mnsc \propto \sigma^{2.73\pm 0.29}$.  This appears to cast doubt on the
notion that NSCs and SMBHs share a common evolutionary path, i.e. that 
they are merely different incarnations of ``central massive object''.  
We speculate that the discrepancy 
between our results and those of \citet{ferrarese06a}, who found 
an $\mnsc-\sigma$ relation closer to $\mnsc \propto \sigma^4$ using 
\textit{observed} velocity dispersions, can 
perhaps be explained if the baryonic mass fraction is not the same for
all early-type galaxies, or follows a different radial profile than the
luminous matter.

Finally, we have compared the observed present-day NSC masses to 
first-order predictions of two proposed NSC formation scenarios, namely
GC infall and feedback-regulated in-situ formation. We find that neither
scenario can provide a reasonable agreement with the observations.

\section*{Acknowledgments}

We would like to thank Elizabeth Wehner for useful suggestions that improved 
the quality of our paper, and Laura Ferrarese for helpful discussions regarding the
measured velocity dispersion of the ACSVCS galaxies.

\bsp

\label{lastpage}

\end{document}